# Transport properties of photonic topological insulators based on microring resonator array

Xiaohui Jiang, Yujie Chen*, Chenxuan Yin, Yanfeng Zhang, Hui Chen, and Siyuan Yu
State Key Laboratory of Optoelectronic Materials and Technologies, School of Electronics and Information Technology, Sun Yat-sen University, Guangzhou, China, 510275
*Corresponding author: chenyj69@mail.sysu.edu.cn

**Abstract**
An array of ring resonators specifically designed can perform as a topological insulator. We conduct simulations using both Tight-Binding Model (TBM) and Transfer Matrix Method (TMM) to analyze the transport properties of such optical structure, verifying the presence of robust topological edge states which is immune to disorder and defect. We have also made a comparison between these two methods, of which results suggesting that TBM is only applicable under weakly-coupling condition while TMM is more rigorous. Finally we compared the structure with common microring array and coupled resonator optical waveguide (CROW) to demonstrate that it has desired transmission properties with wide and flat spectral response.

Topological insulator is a special kind of insulator which has typical insulating energy band structure while a Dirac-cone type of band structure appears in the band gap connecting the valance and conduction band. This peculiar characteristic of topological insulator results in surface (for 3D systems) / edge (for 2D systems) conducting which is protected by topology and thus immune to disorder and defect. If achieved in an optical system, this ideal conducting feature without dissipation and back-scattering can be applied to robust transport of light in photonic integration [1].

The idea of photonic topological insulator (PTI) arose first in applying magneto optic phenomena in photonic crystal [2,3]. Following with that edge state is observed in the realm of microwaves [4,5]. Because of the weak interaction of magneto-optic effects between photons and magnetic field at optical frequencies, it is hard to engineer PTI under magnet [6,7]. In 2011 the first optical topological insulator system was achieved by implementing synthetic gauge (magnetic) field for photons which is much like state of electrons interacting with magnet in quantum Hall effect (QHE) system [8-12]. Here, we conduct simulations to analyze the transport properties of PTI built by microring resonator array, applying both Tight-binding model (TBM) and Transfer matrix method (TMM).

We verified that robust edge states exist based on the transmission spectra and the distribution of light intensity around the optical path in the system. In particular, we compared the transmission spectra between common microring resonator array, coupled resonator optical waveguide (CROW) and PTI built by microring resonator array. We found that PTI presents relatively steady spectral response (flat spectrum) even when defects appear. This is what we desire in optical integration.

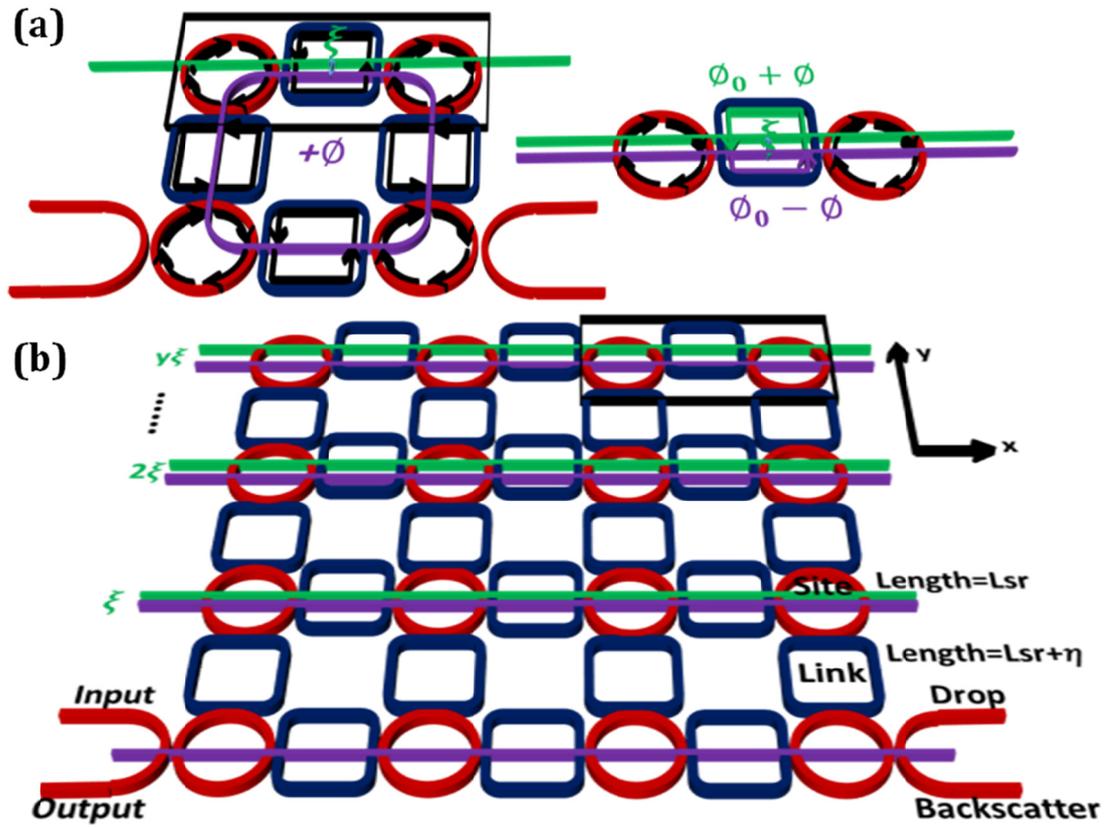

Figure 1: Structure design of microring resonator array. (a) Diagram of single element with synthetic magnetic field. ($\xi$ is the shifted length unit and $\emptyset$ the phase shift caused by it. $\emptyset_0$ is the phase shift caused by half length of link resonator) (b) Resonator array design. ($L_{sr}$ is the length of site resonators and $\eta$ is the extra length of link resonator to provide $\pi$ phase resonance shift compared to site resonator.)

 

    The system is an array of microring resonators with specifically shifted link resonators along *y* direction. Site resonators are equivalent to lattice sites in solid material and photons hop between them with certain probability, just like how electrons behave in solid lattice. Though time symmetry is not broken as it is in QHE system, this optical topological structure does exhibit robust edge states immune to certain disorders despite those induced by backscattering. Tight-binding model is applicable to this system under weakly-coupling condition [13,14]. A more recent study [14] suggested that the array can be made periodic without shift of link resonators along *y* direction but still achieving the same phenomena through strong coupling effect between resonators. TBM is not appropriate here for it is limited to single mode approximation (weak coupling) and thus the more rigorous TMM is applied.

    Synthetic magnetic field is implemented by specifically shifting the link resonators along *y* direction as shown in Fig. 1. The size of site and link resonators and shifted length we used in our simulation is based on silicon nitride ($SiN_x$), which is a promising CMOS-compatible dielectric material applicable to photonic integration for applications in areas with

wavelength range from visible to infrared regions [15]. The length of link resonators is adjusted to have a π phase shift compared to the site resonators so that they do not resonant around the resonance of site resonators thus only act like common light guiding waveguides. The shifted length of link resonators is $y\xi$ depending on the line index. It is chosen to let the magnetic coefficient $\alpha_M$ to be 1/4 so the uniform accumulated phase per plaquette can be $\emptyset = 2\pi \times \alpha_M = \pi/2$ according to Hofstdater spectrum of QHE in order to induce edge state within the PTI system [10,16,17].

Note that discrete Hamiltonian can be used to derive the wavefunction of electrons in crystal lattice. Electrons are bound to the effective periodic potential due to the effect of lattice sites and they have a finite probability to tunnel to other sites. TBM is an approximation that consider only the tunneling between nearest sites which is reasonable since the effect of the nearest is dominant, which is thus suitable to be applied to the weak-coupling microring resonator array [18]. Here TBM Hamiltonian is adjusted to such optical array, treating site resonators as lattice sites. Photons can tunnel to the nearest site with a tunneling rate J and a hopping phase. Given that the lattice is infinite, the Hamiltonian has the form written as [11,20]:

$$H = \sum_{xy} \hat{a}^+_{x,y} \hat{a}_{x,y} - J\left(\sum_{xy} \hat{a}^+_{x+1,y} \hat{a}_{x,y} + \hat{a}^+_{x,y} \hat{a}_{x+1,y} + \hat{a}^+_{x,y+1} \hat{a}_{x,y} + \hat{a}^+_{x,y} \hat{a}_{x,y+1}\right) \quad (1)$$

where $\hat{a}$ is the photon operator in the microring resonator. Hence the tight-binding Hamiltonian of a 5×5 lattice (only considering site resonators) is a 25× 25 matrix:

$$H = \begin{pmatrix} \omega_0 & -J & \cdots & 0 & 0 \\ -J & \omega_0 & & 0 & 0 \\ \vdots & & \ddots & & \vdots \\ 0 & 0 & \cdots & \omega_0 & -Je^{-iy\emptyset} \\ 0 & 0 & & -Je^{iy\emptyset} & \omega_0 \end{pmatrix} \quad (2)$$

where $\omega_0$ is the resonant frequency, $H_{i,j}$ represents the coupling between site $i$ and $j$. $J$ is the coupling rate between adjacent site resonator and $e^{iy\emptyset}$ is the phase induced by synthetic magnetic field.

Implementing energy input and output formalism and couple mode theory, the time evolution of the energy amplitude in the ring resonator can then be expressed as:

$$\frac{da(t)}{dt} = i[H, a(t)] - \frac{a(t)}{\tau} - \sqrt{2k^I_{ex}}\varepsilon_I(t) \quad (3)$$

where the first term on the right side can deduce to the resonant frequency. The second term is the decay rate of the ring, consisting of intrinsic loss ($\kappa_{in}$) and extrinsic loss ($\kappa_{ex}$). The third term is the energy input of the site resonator. Suppose that the input electric field has the form of a plane wave $\varepsilon_I(t) = \varepsilon_I e^{-i\omega t}$, the above equation can be modified into:

$$-i\omega a = i[H, a] - \kappa_{in}a - (\delta_I + \delta_O)\kappa_{ex}a - \sqrt{2\kappa^I_{ex}}\varepsilon_I \quad (4)$$

The transport properties of such optical array thus can be obtained through solving Eq. (4), of which transmission spectra are shown in Fig. 2(a). It can be seen from the transmission spectra at the drop port of the 5×5 microring resonator array that it is split into three parts, corresponding to three different states: short edge state, bulk state, and long edge state, respectively. Short edge state and long edge state refer to as light transporting through the short and long path of the device while bulk state refers to as the randomly distributed electric field intensity within the structure.

We have also verified that the edge states are protected and they present robust transport properties immune to defect by intentionally removing resonator within the array (Fig. 3).

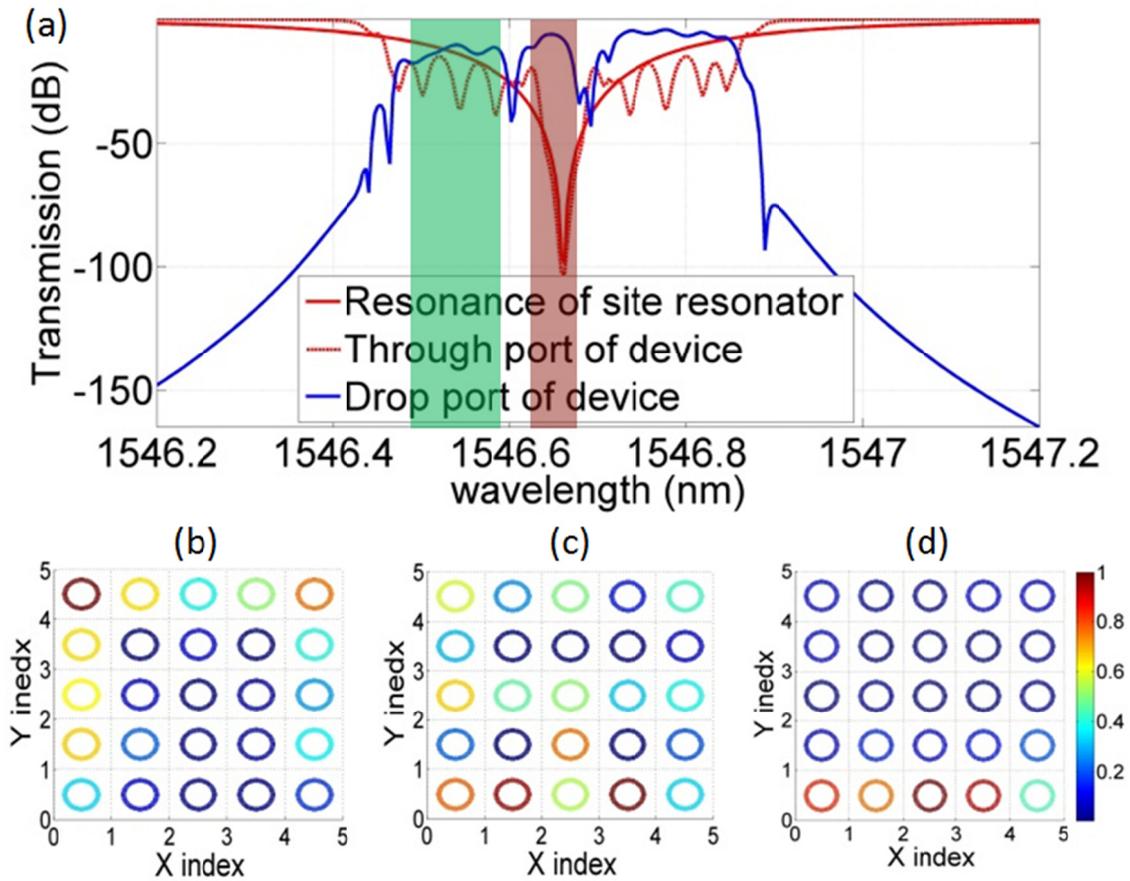

Figure 2: Transmission spectra and electric field power in each ring of 5×5 ring resonator array. (a) Transmission spectra of device. Spectra within color bars represent long edge state (green), bulk state (blue) and short edge state (pink), respectively; (b)-(d) Electric field power (normalized) within each ring corresponding to three states (b, c, d).

From the above analysis, we approve the existence of robust edges with the ability to route around defect and find its way to the output, thus acting as unidirectional and lossless waveguide. It can be noticed that the bandwidth of edge states is larger than the resonance bulk band, suggesting larger tunable ability.

Though robust edge states have been simulated using TBM, it is limited to weakly-coupling condition (i.e. the resonance bandwidth of microring resonator is much narrower than FSR), which satisfies single mode approximation. In the meantime, the exact single mode should be known beforehand and only one mode can be analyzed in TBM. Also the link resonator is simplified to be equivalent to connecting waveguide with certain coupling rate $J$.

While TBM considers mode energy relations between sites, TMM considers amplitude and phase transitions between coupling junctions within the ring array, in which the amplitude and phase transitions between adjacent site and link resonators are taken into consideration.

It regards each ring in the array as a loop, which can be described by transfer matrix approach [19], shown in Fig. 4.

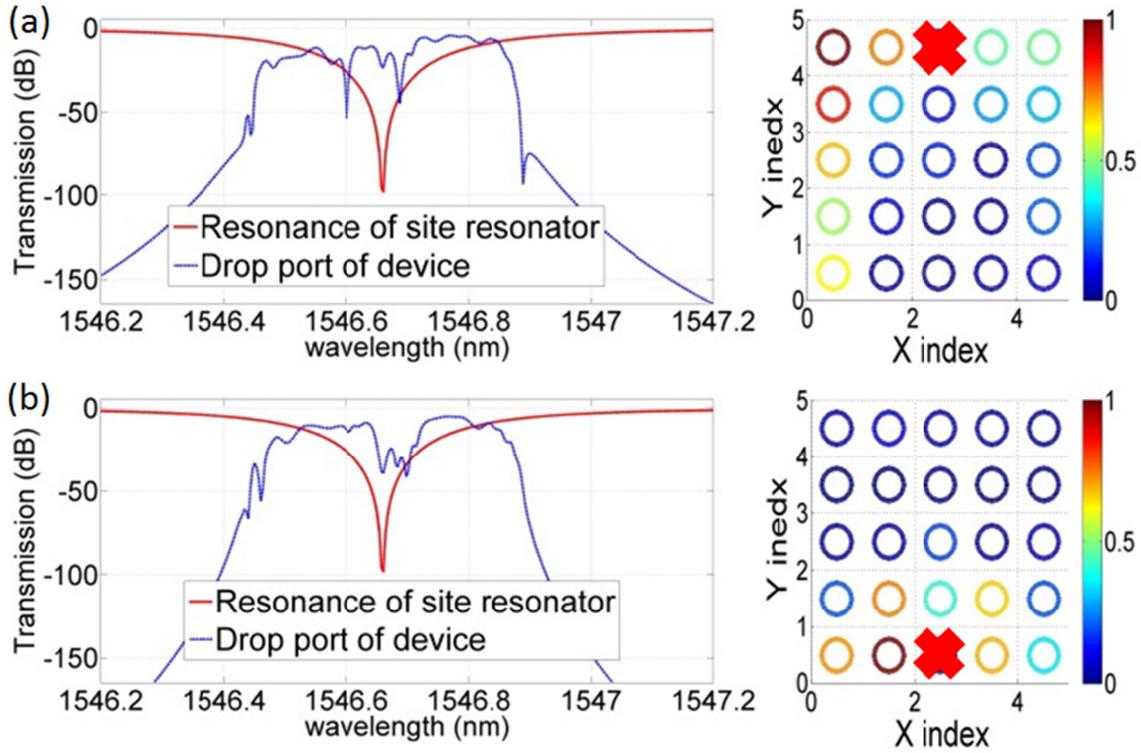

Figure 3: Transmission spectra of device with defect and the corresponding electric field power (normalized) of edge states. A red cross represents a removed ring resonator. (a) Long path. (b) Short path.

Transfer matrix describing the coupling between all nodes *a*, *b*, *c*, and *d* within microring array is built as $M_a$, $M_b$, $M_c$, and $M_d$ [20]. For example, $M_a$ that characterizes the coupling between node *a* to *b* in an 2×2 array shown in Fig. 4(a), is an 8×8 matrix:

$$M_a = \begin{pmatrix} SR & 0 & \cdots & 0 & 0 \\ 0 & LR & & 0 & 0 \\ \vdots & & \ddots & & \vdots \\ 0 & 0 & \cdots & LR*M & 0 \\ 0 & 0 & & 0 & SR \end{pmatrix} \quad (5)$$

where $SR = e^{i\beta L_{sr}/4}e^{i\alpha L_{sr}/4}$ and $LR = e^{i\beta L_{sr}/4}e^{i\alpha L_{sr}/4}$ represent the phase and loss of site and link resonators, respectively. $M = e^{iy\phi}$ corresponds to the accumulated phase induced by synthetic magnetic field. $M_b$, $M_c$, and $M_d$ can be obtained accordingly. Then by implementing electric field amplitude input and output formalism, one can obtain

$$E_a = LME_a + E_I \quad (6)$$

where $M = M_d M_c M_b M_a$ is the transfer matrix of the ring array.

$$L = diag[t, 1, t, \ldots, 1] \quad (7)$$

where *L* is input and output matrix with $L_{ii} = t$ representing input/output through $i$ site ring resonator.

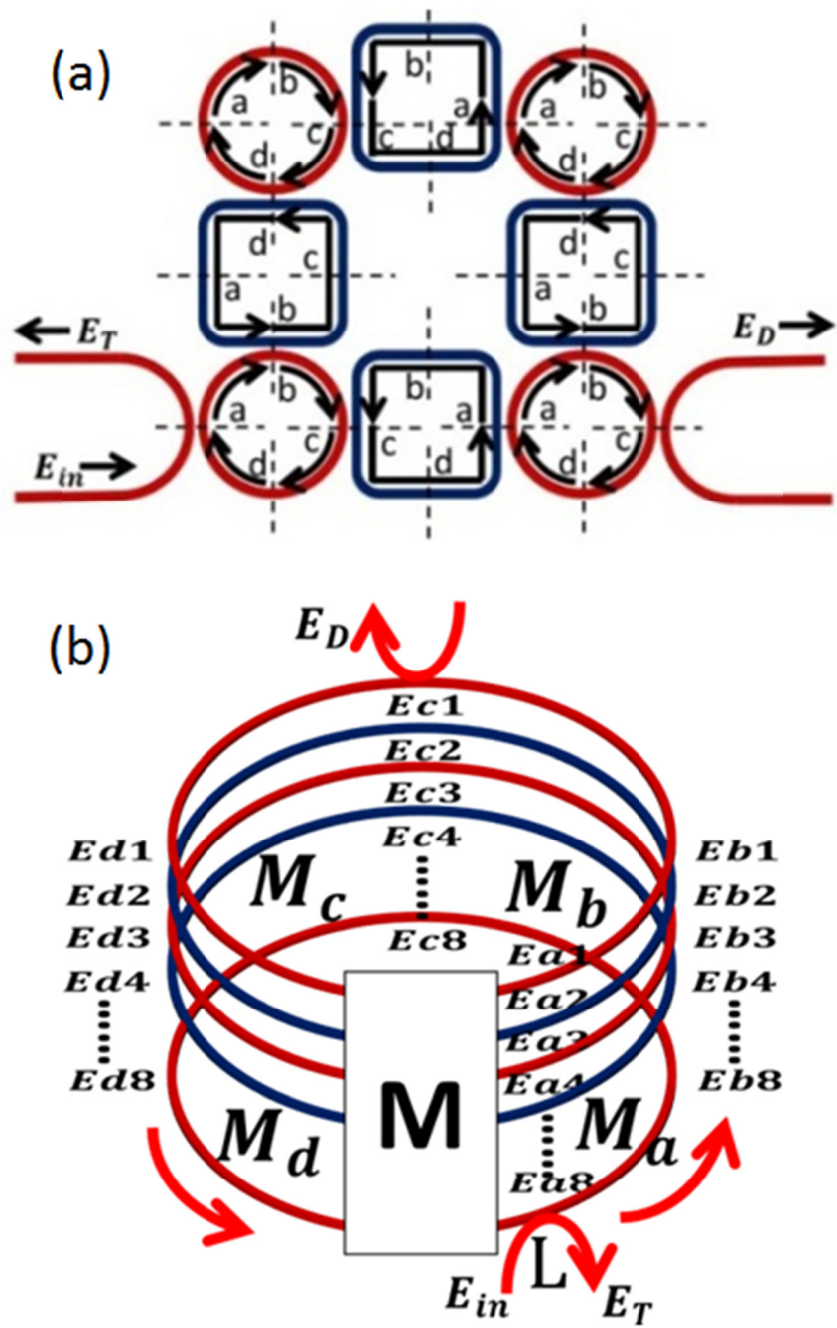

Figure 4: An 2×2 microring resonator array with input and output waveguide and the transfer matrix diagram. (a) Structure of 2×2 lattice with marked electric field nodes a, b, c, and d. (b) Diagram of the straight waveguide matrix net of the coupled ring resonator array.

Simulation results are shown in Fig. 5. The transmission spectrum is much like that calculated via TBM, indicating three different states. Figure 5(b) compares the results between them (marked by the same color bars as in Fig. 2). Two adjacent modes are simulated by TMM and Fig. 5(c) shows the simulated spectra. The mode between two resonances of the site resonator is the mode of the link resonator, since that it has a π phase shift compared to the site resonator. More modes can be simulated through TMM. The accurate solution of the electric field within such microring resonator array can thus be obtained. The robust edge states can be consequently depicted using TMM by removing some ring resonators. Furthermore, TMM is also applicable when the link resonators are not acting just as a common waveguide. It can take into account the resonant properties of link resonators.

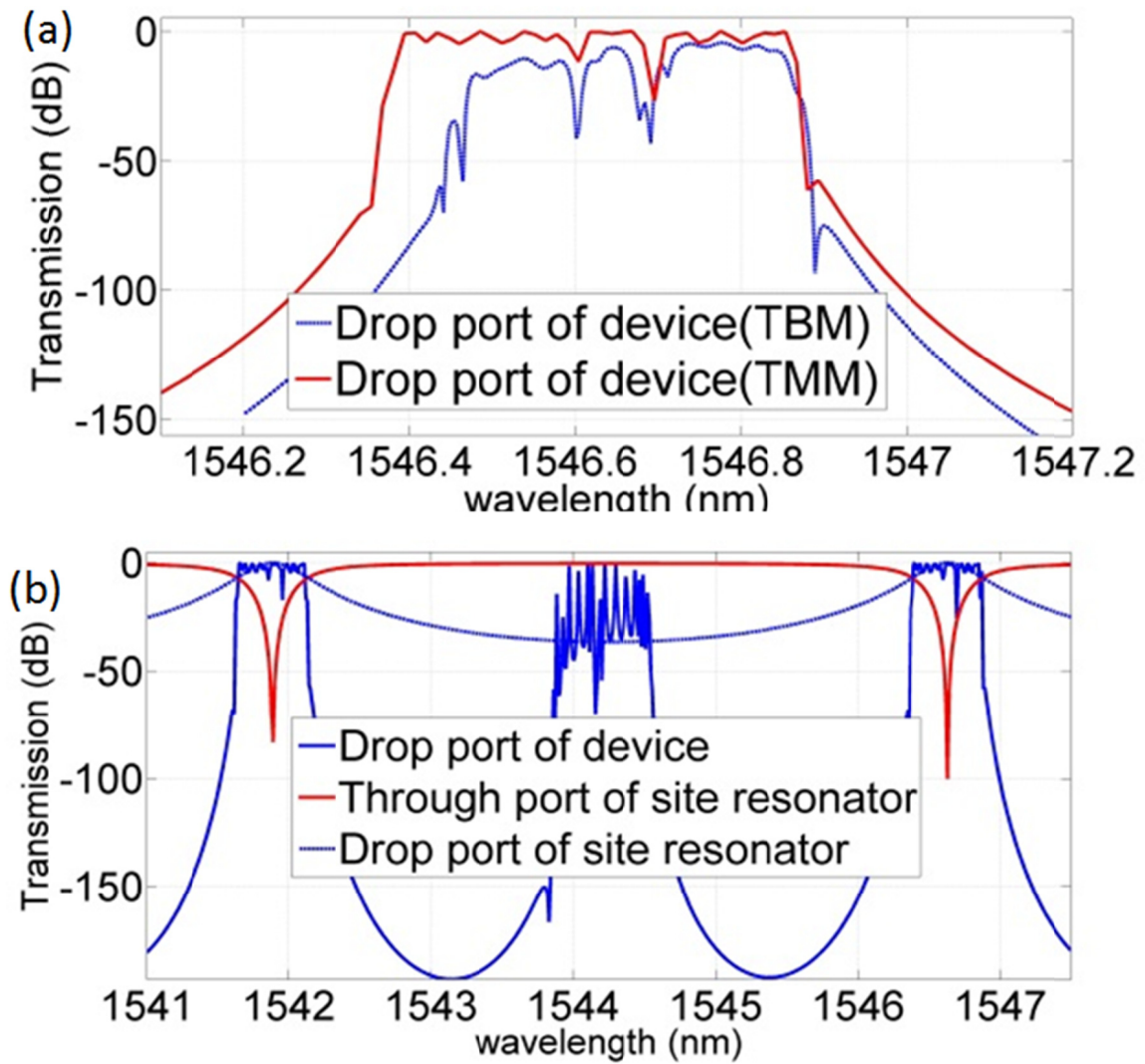

Figure 5: Simulation results of TMM. (a) Comparison between TMB and TMM. (b) Two adjacent modes simulated by TMM.

Now we give further analysis by comparing the transmission spectrum of such PTI with optical microring array without shift in y direction and normal CROW. To obtain the transmission spectra around only one mode, we apply TBM. Typical structures and spectra of common microring resonator array and CROW are shown in Figs. 6(a) and 6(b). The transmission properties of common microring resonator array are not pleasing with narrow bandwidth and fluctuation. Though having wide bandwidth and flat spectral response, CROW will be disabled if any resonator is defected (Fig. 7). In comparison, the spectral response of PTI built by microring resonator array is stable and thus can present robust performance for applications in photonic integration.

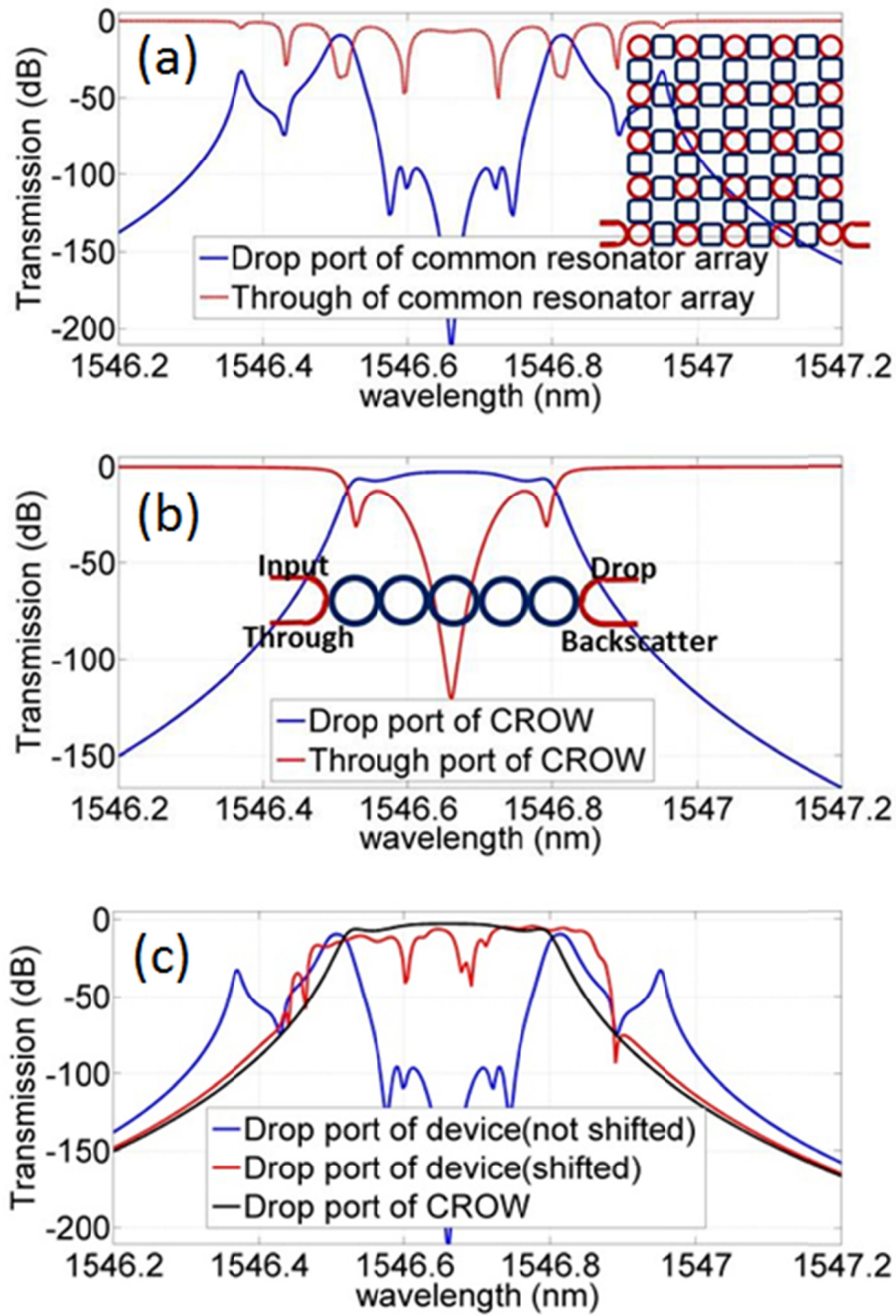

Figure 6: Comparison between transmission spectra of PTI built by microring resonator array, common ring resonator array, and CROW. (a-b) Transmission spectra of common ring resonator array/CROW. (c) Comparison of transmission spectra between different structures.

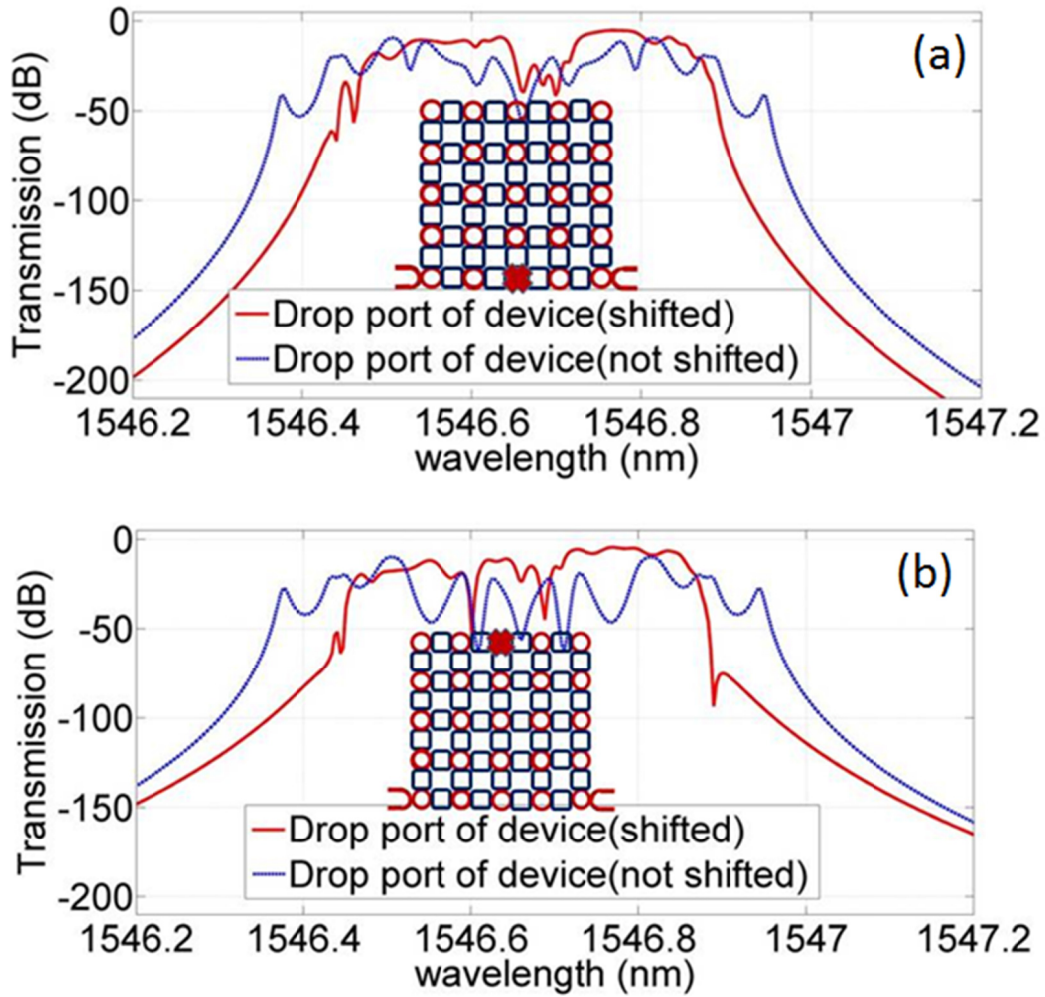

Figure 7: Comparison of transmission spectra when a defect appears. (a-b) Transmission spectra of PTI and common microring array, with the 3[rd]/ 23[th] site resonators defected. (Marked by a red cross)

In conclusion, we have simulated the transport properties of 2D microring resonator array with specific engineered shifts. It successfully achieves the effect as a PTI, with the appearance of edge states which is immune to defects. Through the comparison of transmission spectra for various waveguide structures, we show the delighting properties of PTI built by 2D microring resonator array. Such study thus deepens our understanding of PTI using microring resonator array and may help put forward applications in the areas of integrated photonics utilizing PTI effects.

**Acknowledgment.** This work is supported by National Basic Research Program of China (973 Program) (2014CB340000, 2012CB315702), the Natural Science Foundations of China (61323001, 61490715, 51403244, 11304401), the Natural Science Foundation of Guangdong Province (2014A030313104), Fundamental Research Funds for the Central Universities of China (Sun Yat-sen University: 13lpgy65, 15lgpy04, 15lgzs095, 15lgjc25, 16lgjc16), and Specialized Research Fund for the Doctoral Program of Higher Education of China (20130171120012).